\documentclass[aps,prl,epsfigure,showpacs,twocolumn]{revtex4}
\usepackage{amsmath}
\usepackage{amstext}
\usepackage{latexsym}
\usepackage{psfig}
\usepackage{graphicx}
\usepackage{amsfonts}
\usepackage{bm}
\usepackage{amssymb}

\newcommand{\ket}[1]{\left\vert#1\right\rangle}
\newcommand{\modul}[1]{\left\vert#1\right\vert}

\newcommand{\one}{\mbox{$1 \hspace{-1.0mm}  {\bf l}$}}

\begin{document}

\title{Multi-splitter interaction for entanglement distribution}

\author{H. McAneney, M. Paternostro, and M. S. Kim}

\affiliation{School of Mathematics and Physics, The Queen's University,
Belfast BT7 1NN, United Kingdom}
\date{\today}

\begin{abstract}
In protocols of distributed quantum  information processing, a
network of bilateral entanglement is a key resource for efficient
communication and computation. We propose a model, efficient both
in finite and infinite Hilbert spaces, that performs entanglement
distribution among the elements of a network without local
control. In the establishment of entangled channels, our setup
requires only the proper preparation of a single elected element. We suggest a setup of
electromechanical systems to implement our proposal.
\end{abstract}

\maketitle
The role of entanglement in delocalized architectures of a device
for quantum information processing (QIP) has been investigated
under many aspects~\cite{varie0}. Entanglement between distant
sites of a distributed register is a fundamental
requisite to optimize communication protocols and perform
efficient quantum computation~\cite{varie}. In this context, an
{\it entanglement distributor} creates an entangled network of
the elements of a register that, otherwise, have no direct
 reciprocal interaction. The efficiency of the distributor can be quantified
 by the number of elements which are
entangled per single use of the distributor or by the
amount of entanglement shared by any two of them.
Thus, the choice of the most appropriate design of
the distributor is a problem-dependent issue with no general recipe. An
interesting configuration for this problem
is a star-shaped system in which an elected element
interacts simultaneously with
many other independent subsystems~\cite{huttonbose}.

In this paper, we propose a model that acts as an efficient entanglement
distributor. An important feature of our proposal is that no 
local control on the dynamic of the participating systems 
is required once the interactions are set. We only need the pre-engineering of 
the network and a proper control of the interaction time. This is an advantage exploitable
in those situations (frequent in solid-state physics) where
single-element addressing is hard or impossible. The interaction 
we suggest acts on a multipartite bosonic network
whose evolution can be tracked analytically both in the discrete and the continuous variable (CV) case.

Despite our proposal being naturally described using the quantum
optics language, we show that our model is general enough to find
interesting applications in solid-state physics. We sketch a
system of coupled electromechanical oscillators to embody our
model. Similar setups have recently found applications in the entanglement-transmission problem~\cite{plenio}.

\begin{figure} [t]
{{\bf(a)}}\hspace*{4.0cm}{\bf(b)}
\centerline{\psfig{figure=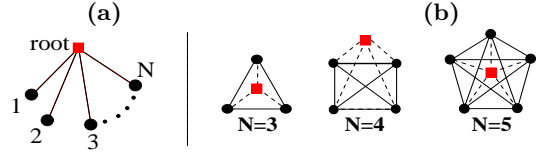,width=7.0cm,height=1.6cm}}
\caption{{\bf (a)}: The interaction configuration in Eq.~(\ref{interazione}). 
Each edge represents an {\it interaction}. {\bf (b)}: Complete entanglement graph generated by Eq.~(\ref{interazione}). Solid and dashed edges represent entanglement.} 
\label{graphs}
\end{figure}

{\it The model -} We consider $N$ bosons (or {\it modes}) $b_j$
($j=1,..,N$) described by the annihilation (creation) operators
$\hat{b}_{j}$ ($\hat{b}^{\dagger}_{j}$) and an additional mode,
labeled $a$, which we call the ${\it root}$. The interaction
configuration is sketched in Fig.~\ref{graphs} {\bf (a)} and
consists of the resonant couplings of the root to each $b_{j}$.
The satellite elements $b_{j}$ do not mutually interact. In the
interaction picture, we consider the Hamiltonian
\begin{equation}
\label{interazione}
\hat{H}_{I}=\sum^{N}_{j=1}\mathcal{G}_{j}\hat{a}^{\dag}\hat{b}_{j}+h.c.\hskip0.5cm(\hbar=1)
\end{equation}
with $\mathcal{G}_j$ real and time-independent couplings. 
For $N=1$, $\hat{U}(\tau)=e^{-i
\hat{H}_{I}\tau}$ is similar to a beam-splitter (BS) superimposing
mode $a$ to $b_{1}$. We analyze the characteristics of the many-body dynamics corresponding to $N>1$.
After a lengthy calculation based on Lie algebra we find that
$\hat{U}(\tau)$ can be decomposed as
\begin{equation}
\label{evoluzione}
\begin{aligned}
\hat{U}(\tau)&=\left[\otimes^{N-1}_{j=1} \hat{R}_{b_{j}}
\hat{B}_{b_{j+1}b_{j}} (\varepsilon_{j},0)\right] \hat{B}_{b_{N}a}(\vartheta_{N}\tau,-{\pi}/{2})\\
&\otimes\left[\otimes^{N-1}_{j=1}\hat{R}_{b_{N-j}}\hat{B}_{b_{N-j+1}b_{N-j}}(\varepsilon_{N-j},0)\right],
\end{aligned}
\end{equation}
where $\vartheta^2_{k}=\sum^{k}_{j=1}{\cal G}^{2}_{j}$.
$\hat{R}_{b_{j}}=e^{i\pi\hat{b}^{\dagger}_{j}\hat{b}_{j}}$ is a $\pi$-phase shifter for mode $b_{j}$, $\varepsilon_{j}=\cos^{-1}({\cal G}_{j+1}/\vartheta_{j+1})$ and $\hat{B}_{ab}(v,\varphi)=e^{[v(\hat{a}^{\dagger}\hat{b}e^{i\varphi}-\hat{a}\hat{b}^{\dagger}e^{-i\varphi})]}$ denotes a BS operator. {\it This decomposition is extremely useful as it shows that the dynamic can be interpreted as the action of a setup of optical elements on $N+1$ bosons}. 
Eq.~(\ref{evoluzione}) describes
how the root gains information from $b_{j}$'s via the
interaction $\hat{B}_{b_{N}a}$ as well as the distribution of any
information initially in $a$ to the satellites. The form of Eq.~(\ref{evoluzione})
 reveals that, if $b_{j}$'s are all prepared in rotationally-invariant states (such as $|0\rangle$ or thermal states), the transformations prior to $\hat{B}_{b_{N},a}$
do not contribute to the entanglement
dynamics~\cite{myungBS}. By properly setting the $\varepsilon_{j}$, the
evolution of the network can be made equivalent to an array of
BS's which sequentially superimpose $a$ to $b_{j}$ modes. If $a$ is in a superposition of $\ket{0}$ and a coherent state, $b_{j}$'s being in the vacuum state, we generate an $(N+1)$-mode GHZ state useful for secret
sharing~\cite{hillery}. The entire Eq.~(\ref{evoluzione}) must be
considered if we initially prepare one or more satellite modes in a
coherent or a non-classical state. 

The model described by Eq.~(\ref{interazione}) realizes various
interference patterns in the equivalent all-optical setting  
allowing for different tasks.  For instance, if ${\cal G}_{j}={\cal G}$
($\forall{j}$), $\hat{H}_{I}$ describes an effective $XY$-coupling
suitable for $1\rightarrow{N}$ phase-covariant 
cloning~\cite{clone}. As another example, let us take $N=2$ so that
Eq.~(\ref{evoluzione}) reduces to
$\hat{U}(\tau)=\hat{R}_{b_{1}}\hat{B}_{b_{2}b_{1}}(\varepsilon_{1},0)
\hat{B}_{b_{2}a}(\sqrt{2}{\cal G}\tau,-\pi/2)
\hat{R}_{b_{1}}\hat{B}_{b_{2}b_{1}}(\varepsilon_{1},0)$  
with $\varepsilon_{1}=\pi/4$. We assume that mode $b_{1}$ is initially
prepared in the 
single-excitation state $\ket{1}_{b_{1}}$, $b_{2}$ and $a$ being in
the vacuum 
(the investigation can be generalized
to the case of $b_{1}$ being prepared in a coherent state). It is
easily seen from our decomposition that at $\sqrt{2}{\cal
  G}\tau={\pi}$, the initial state 
is transferred to mode $b_{2}$ with unit probability (while the maximum
probability of finding the initial state in $a$ is only $1/2$). This
analysis shows that   
perfect quantum state transfer from $b_{1}$ to $b_{2}$ can be
performed through mode $a$. In fact, when $\sqrt{2}{\cal G}\tau=n\pi$
($n=0,1,\cdots$), it is interesting to note that our model is
equivalent to a Mach-Zehnder 
interferometer with a $n\pi$ 
phase-shift in the path of mode $b_{2}$, which is
obvious from our 
decomposition in Eq.~(\ref{evoluzione}). 

{\it Single excitation case - } Consider $a$ initially prepared in $\ket{1}_{a}$, $b_{j}$'s being in
$\otimes^{N}_{j=1}\ket{0}_{j}$.
The dynamics is captured by
considering a fictitious collective mode of its annihilation
operator
$\hat{c}=\sum^{N}_{j=1}{\cal G}_{j}\hat{b}_{j}/\vartheta^{}_{N}$. Thus,
$\hat{U}(\tau)\ket{10..0}_{ab_{1}..b_{N}}=\cos(\vartheta_{N}\tau)
\ket{1\underline{0}}_{ac}-i\sin(\vartheta_{N}\tau)\ket{0\underline{1}}_{ac}$
with $\ket{\underline{1}}_{c}=
\hat{c}^{\dag}\ket{0..0}_{b_{1}..b_{N}}=\sum_{j}({\cal G}_{j}/\vartheta^{}_{N})\ket{0..1..0}_{b_{1}..b_{j}..b_{N}}$.
This state can be pictorially described
by complete {\it entanglement graphs} as those shown in
Fig.~\ref{graphs} {\bf (b)}. There, solid or dashed edges represent entanglement. 

In the basis $\{\ket{00},\ket{01},\ket{10},\ket{11}\}_{b_{i}b_{j}}$, the reduced 
density matrix of the generic pair $b_{i},\,b_{j}$ ($\forall\,i,j$) reads
\begin{equation}
\label{matrixqubit}
\bm{\rho}_{{i}{j}}=
\begin{pmatrix}
1-(G^{2}_{iN}+G^{2}_{jN})&0&0&0\\
0&G^{2}_{jN}&G_{iN}G_{jN}&0\\
0&G_{iN}G_{jN}&G^{2}_{iN}&0\\
0&0&0&0
\end{pmatrix},
\end{equation}
where $i<j$ and $G_{jN}={\cal G}_{j}\sin(\vartheta_{N}\tau)/\vartheta_{N}$. The entanglement
of this mixed state can be quantified by the {\it concurrence}
$C_{N}=\max\left\{0,{\alpha_{1}}-{\alpha_{2}}-{\alpha_{3}}-{\alpha_{4}}\right\}
$~\cite{concurrence}. Here, $\alpha_{i}$'s are the square
roots of the eigenvalues (in non-increasing order) of
$\bm{\rho}_{ij}(\bm{\sigma}_{y}\otimes\bm{\sigma}_{y})\bm{\rho}^{*}_{ij}
(\bm{\sigma}_{y}\otimes\bm{\sigma}_{y})$ with $\bm{\rho}^{*}_{ij}$
the complex conjugate of $\bm{\rho}_{ij}$ and $\bm{\sigma}_{y}$
the $y$-Pauli matrix. We get $C_{N}=\max\{0,2G_{iN}G_{jN}\}$.
For later purposes, it is also useful to consider the entanglement
measure  based on {\it negativity of partial transposition}
(NPT)~\cite{npt}. NPT is a necessary and sufficient condition
for entanglement of any bipartite qubit state~\cite{npt}. The
corresponding entanglement measure is defined as
$NPT_{N}=\max\{0,-2\lambda^{-}\}$ with $\lambda^{-}$ the negative
eigenvalue of $\bm{\rho}^{T_{j}}_{ij}$ which is the partial
transposition of $\bm{\rho}_{ij}$ with respect to $b_{j}$. We find
${NPT}_{N}=\max\{0,{[(1-G^{2}_{iN}-G^{2}_{jN})^2+4G^{2}_{iN}G^{2}_{jN}]^{1/2}}-(1-G^{2}_{iN}-G^{2}_{jN})\}$.
$C_{N}$ and $NPT_{N}$ are optimized when
$\vartheta_{N}\tau=(2k+1)\pi/2$ ($k\in{\mathbb Z}$). Using this
condition as a constraint in the Lagrange's method  of
indeterminate multipliers, we find that $C_{N}$ and $NPT_{N}$ are
maximized for the uniform set of couplings ${\cal G}_{j}={\cal G}$
($\forall{j}$). In this case we get $C_{N,max}=2/N$ and
$NPT_{N,max}=\{[{4+(N-2)^2}]^{1/2}-(N-2)\}/N$. $2/N$ is the
upper bound for the bipartite entanglement in an $N$-party
system~\cite{buzekimoto}. Thus, Eq.~(\ref{interazione}) is optimal under the point of view of
pairwise entanglement distribution. For equal ${\cal G}_{j}$, the
$\bm{\rho}_{ij}$ are all equal and we have $\ket{10..0}_{ab_{1}..b_{N}}{\rightarrow}\cos(\vartheta_{N}\tau)\ket{1,0..0}_{ab_{1}..b_{N}}
-i\sin(\vartheta_{N}\tau)\ket{0,W_{N}}_{ab_{1}..b_{N}}$. We have introduced the $N$-particle $W$-state
$\ket{W_{N}}_{b_{1}..b_{N}}=N^{-1/2}\sum_{j}\ket{0..1..0}_{b_{1}..b_{j}..b_{N}}$ which is the state achieving $C_{N,max}$\cite{buzekimoto}. Thus, the maximum concurrence between any pair of $b_{j}$'s is found
when the root is separable from the rest of the network. The
corresponding graph is obtained by deleting the
dashed edges in Fig.~\ref{graphs} {\bf (b)}, the satellite
elements forming complete and permutation-invariant entanglement graphs. 
The system periodically evolves from a separable state to a configuration where the root is factorized from the rest of the network
(which is in $\ket{W_{N}}_{b_{1}..b_{N}}$).
In between, an ($N+1$)-partite entangled state is obtained.

Recently, a configuration of many spin-$1/2$ systems
analogous to Eq.~(\ref{interazione}) has been
proposed~\cite{huttonbose}. The one-excitation case we have
considered allows for a comparison between the two situations,
 both achieving $C_{N,max}=2/N$. In our model
the bosonic nature of the register allows for this result without
local control on the satellite elements or the root. In
ref.~\cite{huttonbose}, on the other hand, this is obtained 
by using an additional
magnetic interaction and through the measurement of the state of the root.

In order to further characterize our entanglement
distributor, we compare $\ket{W_{N}}_{b_{1}..b_{N}}$ to the class
of {\it cluster states}. These are known to be useful and genuine
multipartite entangled states~\cite{cluster}, inequivalent to 
$\ket{W_{N}}$ for any $N$. While there are always 
proper local measurements on a subset of a cluster that allow for
the deterministic extraction of a pure Bell state, this is not
the case for a $W$-state. However, the quantum correlations in a
cluster are encoded in the system as a whole and any pairwise
entanglement (obtained by tracing out the rest of the cluster) is
zero. This is a drawback in those situations where bipartite
entanglement is required but the physical system is such that the realization of a measurement pattern is made difficult by the problems related to single-element addressing. Finally, the entanglement of $\ket{W_{N}}$ is {\it persistent} as $N-1$ projective measurements are required in order to disentangle the elements of the register.
For the problem we address here, our analysis shows
that Eq.~(\ref{interazione}) is a suitable and exploitable model.

{\it CV case - }Considering only the case of a single excitation
in the root restricts the possibilities offered by the bosonic
nature of our register. In ref.~\cite{myungBS} it is shown that a
non-classical input is a fundamental pre-requisite for the
entanglement of the outputs of a beam-splitter. The same is true
in our case because of the analogy between a BS and
Eq.~(\ref{evoluzione}). On the other hand, necessary and sufficient conditions
for the entanglement are known  and entanglement can
be quantitatively determined only for the class of
two-mode CV Gaussian states~\cite{simon,myungmunro}. In virtue of these 
considerations and because of the Gaussian-preserving nature of the linear
operations in~(\ref{evoluzione}), only Gaussian states will be considered here.

A powerful tool in the analysis of $N$-mode CV systems is given by the
variance matrix $\bf{V}$, defined (after unitary
displacements) as $V_{\alpha\beta}=\langle\{\hat{x}_{\alpha},\hat{x}_{\beta}\}
\rangle\,(\alpha,\beta=1,..,N)$. Here,
${\bf\hat{x}}=(\hat{q}_{1},\hat{p}_{1},..,\hat{q}_{N},\hat{p}_{N})^{T}$
is the vector of the quadratures. A Gaussian state is fully characterized by the
knowledge of just the first and second moments of ${\bf\hat{x}}$ and, 
in order to characterize the state of our $N+1$ modes, we need
to find the variance matrix of their joint state after
$\hat{U}(\tau)$. In phase-space, the action of $\hat{U}(\tau)$ 
is such that ${\bf V'}={\cal T}^{T}{\bf V}{\cal T}$ becomes the new
variance matrix. Here, ${\cal T}$ is the $2(N+1)\times{2(N+1)}$
unitary matrix (found using Eq.~(\ref{evoluzione}))
\begin{equation}
\label{trasformazione}
{\cal  T}=
\begin{pmatrix}
\cos{(\vartheta_{N}\tau)}\one_{2}&A_{1}{\bm\sigma}_{y}&A_{2}{\bm{\sigma}}_{y}&\cdots&A_{N}{\bm\sigma}_{y}\\
A_{1}{\bm\sigma}_{y}&D_{11}\one_{2}&D_{12}\one_{2}&\cdots&D_{1N}\one_{2}\\
\vdots&\cdots&\cdots&\ddots&\vdots\\
A_{N}{\bm\sigma}_{y}&D_{N1}\one_{2}&D_{N2}\one_{2}&\cdots&D_{NN}\one_{2}&
\end{pmatrix},
\end{equation}
where $\one_{2}$ is the 2$\times$2 identity matrix,
$A_{n}=-iG_{nN}$ and
$D_{nm}=\delta_{nm}+[{\cal G}_{n}{\cal G}_{m}(\cos(\vartheta_{N}\tau)-1)/\vartheta^2_{N}]$. $\delta_{nm}$ denotes the Kronecker symbol.

\begin{figure}[b]
{\bf (a)}\hskip4.0cm{\bf (b)}
\centerline{\psfig{figure=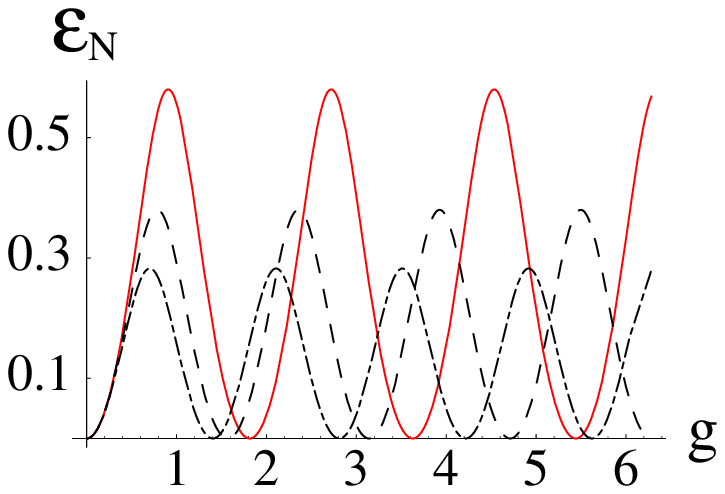,width=4.5cm,height=2.8cm}
\psfig{figure=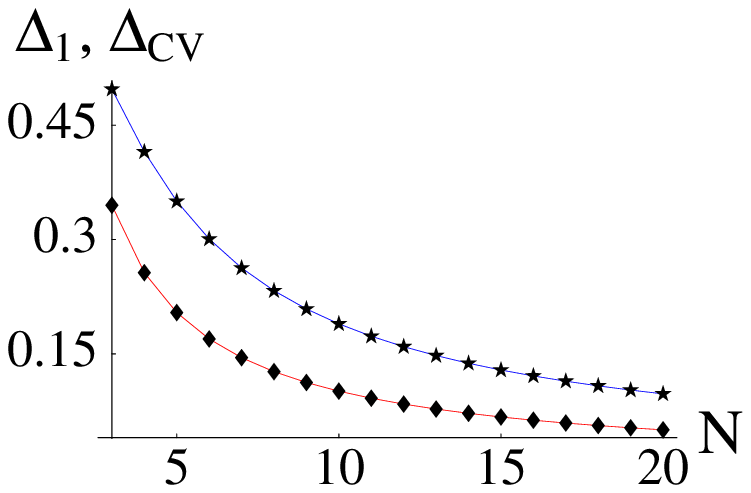,width=4.2cm,height=2.8cm}}
\caption{{\bf (a)}: ${\cal E}_{N}$ against the dimensionless time $g={\cal G}\tau$ for $N=3$ (solid line), $N=4$ (dashed line) and $N=5$
(dot-dashed). The squeezing of the initial root
state is $r=0.8$. {\bf (b)}: Relative entanglement differences
$\Delta_{1}$ ($\bigstar$) and $\Delta_{CV}$ ($\blacklozenge$) against $N$.} 
\label{entCV}
\end{figure}

For simplicity, we take ${\cal G}_{j}={\cal G}$ ($\forall{j}$), $b_{j}$'s
being in the vacuum state (variance matrix ${\bf
V}_{b_{1}..b_{N}}=\oplus^{N}_{j=1}\one_{2b_{j}}$). The root is
prepared in a squeezed state (squeezing parameter $r$) 
which is the most natural non-classical Gaussian state~\cite{myungBS}. The
initial variance matrix of the system is ${\bf V}_{a}\oplus{\bf
V}_{b_{1}..b_{N}}$ with ${\bf V}_{a}\equiv{e}^{-r\bm{\sigma}_{z}}$ the variance matrix of $a$ and $\bm{\sigma}_{z}$ is the $z-$Pauli matrix.
By tracing all the modes but $b_{i}$ and $b_{j}$ 
we get
\begin{equation}
\label{VMB}
{\bf V'}_{b_{i}b_{j}}=
\begin{pmatrix}
{\bf L}_{N}&{\bf C}_{N}\\
{\bf C}_{N}&{\bf L}_{N}
\end{pmatrix}.
\end{equation}
Here, ${\bf L}_{N}\!=\!\mbox{diag}(n_{N},m_{N}),\,{\bf C}_{N}\!=
\!\mbox{diag}(c_{N},d_{N})$ with $n_{N}=1+c_{N}$, $m_{N}=1+d_{N}$,
$c_{N}=-e^{r}d_{N}=(e^{r}-1)\sin^{2}(\vartheta_{N}\tau)/N$. No
dependence on the indices $i,j$ exists so that Eq.~(\ref{VMB}) is the
same for any pair.~${\bf V'}_{b_{i}b_{j}}$ has a form which allows us
to quantify the bipartite entanglement. Indeed, for a variance matrix as
Eq.~(\ref{VMB}), the NPT entanglement measure is given by ${\cal
E}_{N}=\max\{0,(\delta_{1}\delta_{2})^{-1}-1\}$ with
$\delta_{1}=n_{N}-\modul{c_{N}}$ and
$\delta_{2}=m_{N}-\modul{d_{N}}$~\cite{myungmunro}. 
We have
\begin{equation}
\label{ent}
{\cal E}_{N}=\max\left\{0,\frac{2(1-e^{-r})\sin^{2}(\vartheta_{N}\tau)}{N-2(1-e^{-r})\sin^{2}(\vartheta_{N}\tau)}\right\},
\end{equation} which is plotted in Fig.~\ref{entCV}~{\bf (a)} against the effective coupling $g={\cal G}\tau$.
${\cal E}_{N}$ diminishes as $N$ increases and, for fixed
values of $r$, is maximized at
$\vartheta_{N}\tau=(2k+1)\pi/2$ ($k\in{\mathbb Z}$). In
Fig.~\ref{entCV} {\bf (a)} only ${\cal
E}_{N\ge{3}}$ is shown as $N=2$ requires some comments. For this
particular case, by generalizing the results of the analysis in refs.~\cite{myungBS,referee}, 
we expect the evolved state of modes $b_{1},\,b_{2}$ to be locally equivalent to a two-mode squeezed vacuum. 
This result is crucially dependent on the fact that, from Eq.~(\ref{evoluzione}) for $N=2$ and $\vartheta_{2}\tau=\pi/2$, the interaction between the satellite modes is an effective $50:50$ BS. This allows us to decompose the variance matrix of the resulting two-mode state as ${\bf V'}_{b_{1}b_{2}}={O}(\frac{-r}{4}){\bf S}_{b_{1}b_{2}}(\frac{r}{4}){O}(\frac{-r}{4})$. Here $O(\frac{-r}{4})=\otimes^{N}_{j=1}{\bf S}_{b_{j}}(\frac{-r}{4})$ with ${S}_{b_{j}}$ the single-mode squeezing transformation (which does not modify the entanglement structure) 
and ${\bf S}_{b_{1}b_{2}}(\frac{r}{4})$ the variance matrix of a two-mode
squeezed vacuum~\cite{myungmunro}. The state is pure which implies the separability of $a$ from $b_{1}+b_{2}$. 
By studying the purity ${\cal P}_{b_{1}b_{2}}={[\det{\bf
V'}_{b_{1}b_{2}}]}^{-\frac{1}{2}}$~\cite{myungmunro}, we find that
its period is one-half the period of ${\cal E}_{2}$. That is, the
$b_{1}+b_{2}$ state is pure not only when ${\cal E}_{2}$ is
maximum (at $\tau_{odd}=(2k+1)\pi/(2\sqrt{2}{\cal G})$) but also at
$\tau_{even}=k\pi/(\sqrt{2}{\cal G})$, which corresponds to ${\cal E}_{2}=0$. 
 By using the biseparability condition of a
boson from a group of $N$ others~\cite{ww}, we have also checked that at
$\tau_{even}$ no entanglement is found  between $a$ and $b_{1}+b_{2}$. The state is fully separable.

By enlarging the network to $N\ge{3}$, we notice that the first interaction in Eq.~(\ref{evoluzione}) between two satellite modes is a BS of its $\varepsilon_{N-1}=1/\sqrt{N}$, which is no longer a $50:50$ BS. This stops the possibility of getting a state locally equivalent to a two-mode squeezed state \cite{spiegazione} and the structure of the multi-mode entangled state becomes much more complicated than the simple case of $N=2$. In particular, the state of any pair
$(b_{i},b_{j})$ is pure just at $g_{even}$ but no longer when
${\cal E}_{N}$ is maximum. Thus, quantum correlations are shared
between $b_{j}$'s at $\tau_{odd}=(2k+1)\pi/(2\sqrt{N}{\cal G})$ but not
between the root and them. The entanglement configuration {\it
alternates} between a fully separable state and a many-body
entangled state of just $b_{j}$'s, passing by a configuration
in which entanglement is shared with $a$. The picture given by
the graphs in Fig.~\ref{graphs} {\bf (b)} is still valid.

We now look at the effect of increasing $N$ on the properties
on the entanglement distribution. We consider the quantities
$\Delta_{1}=1-(NPT_{N+1,max}/NPT_{N,max})$ and
$\Delta_{CV}=1-({\cal E}_{N+1,max}/{\cal E}_{N,max})$ which
measure the {\it relative loss in pairwise entanglement} if the
network is enlarged by one element. Fig.~\ref{entCV} {\bf
(b)} shows that at a fixed $r$, $\Delta_{1}$ and $\Delta_{CV}$
decrease with $N$ ($3\le{N}\le{20}$). The distribution process is
only weakly affected and the entanglement is still spread through
the network. In passing, it is interesting to stress the qualitative robustness 
of the distributed entanglement in the CV case as compared to the discrete one,
an issue which, in a different context, has also been noticed in \cite{wvc}. 

{\it Possible setups - }We briefly mention that, to embody
Eq.~(\ref{interazione}), we can use the interaction of a linearly
polarized optical bus with $N$ ensembles of cold atoms (confined
in vapor cells), providing the Hamiltonian
$\hat{H}_{le}=\kappa\hat{p}_{ph}\sum^{N}_{i=1}\hat{p}_{ei}$
($\kappa$ is a coupling rate). Here, $\hat{p}_{ph}$
($\hat{p}_{ei}$) is the momentum operator of the bus ($i^{th}$
atomic ensemble) whose wavelength is assumed to be much
larger than the dimensions of the ensembles and their
separations~\cite{polzik}. $\hat{H}_{le}$ holds within the
Stokes-vector formalism for the bus and the Holstein-Primakoff 
transformation mapping collective states of an ensemble
into a fictitious boson. By discarding rapidly-oscillating terms, $\hat{H}_{le}\rightarrow\hat{H}_{I}$.

Stimulating opportunities come from micro  and
nano-electromechanical systems (MEMS and NEMS), {\it i.e.}
electrically controlled mechanical oscillators (or {\it
cantilevers}) whose dimensions are in the range from $10^{-9}$ to $10^{-6}\,m$. 
Doubly clamped cantilevers with
fundamental mode frequency in the range of $[10^7-10^{9}]\,Hz$
have been fabricated and mutually coupled~\cite{buks}. They are useful to
study Heisenberg-limited measurements~\cite{knobel} and
entanglement~\cite{plenio,armour}. There are theoretical proposals for
ground-cooling and squeezing of NEMS mode~\cite{zoller}.
The preparation of phonon-number states and the tomography
of a vibrational mode have also been addressed~\cite{zoller}.

We consider $N$ classical oscillators coupled  via spring-forces
to a central one, the analogue of our root. Within Hooke's law,
the energy of the system is ${\cal
H}=(\omega/2)(q^2_{a}+p^{2}_{a})+(\omega/2)\sum_{j}[q^{2}_{j}+p^{2}_{j}+{\cal
K}_{j}(q_{j}-q_{a})^2]$, where the ${\cal K}_{j}$'s are the
coupling factors, $(q_{j},\,p_{j})$ are proper canonical
variables and $\omega$ is the frequency of the oscillators (equal
for all). Each ${\cal K}_{j}$ is controlled via voltage biases
between the cantilevers. Each bias creates a potential that
changes with the capacitance between two oscillators.
Eq.~(\ref{interazione}) is then found in a second-quantization
picture and within the rotating wave approximation (used for ${\cal
K}_{i}\simeq0.1{\omega}$).
The oscillators can be built via  photolitography of gold on
silicon substrates~\cite{buks}. In our case, planar grids of a
few cantilevers face each other in pairs, surrounding the root.
The coupling of the cantilevers to the phononic modes of the
substrate is the main source of decoherence. However, oscillators
with quality factors $Q\simeq{10}^{4}$ and $\omega\simeq{10}\,MHz$
(coherence times $\simeq1$ msec) allow now for a good number of
coherent operations. The reconstruction of ${\bf V}'_{b_{i}b_{j}}$ is challenging here. However, a single-electron-transistor (SET) capacitively
coupled to the cantilevers can be used~\cite{armour}. Exploiting
the changes of the coupling capacitances (which depend on the
instantaneous position of the oscillators), a
SET acts as a displacement-to-current transducer with
displacement sensitivity $\simeq10^{-16}\,m/\sqrt{{Hz}}$.
Stroboscopic techniques to infer ${\bf V}'_{b_{i}b_{j}}$ could
then be used~\cite{plenio}.


{\it Remarks - }We have characterized a many-body interaction
that,  through just global interaction with a seeding system,
distributes entanglement in a network of local processors. The
dynamics is described by linear operations and the model is
flexible enough to allow for different interference patterns by
pre-engineering the couplings and the initial state. We have
shown how symmetric bipartite entangled states are generated both
in the discrete and CV case. To embody our model, we have
described a setup of coupled cantilevers that offers nice
perspectives in the study of entanglement distributors for QIP.


\acknowledgments

{\it Acknowledgements}- We thank Profs. P. L. Knight and J. Lee for discussions. This work has been supported by the UK
EPSRC, DEL and IRCEP.


\begin{thebibliography}{99}

\bibitem{varie0} D. Gottesman and I.L. Chuang, {\sl Nature} (London) {\bf 402}, 390 (1999);
L.-M. Duan {\it et al.}, {\sl ibid.} {\bf 414}, 413 (2001).

\bibitem{varie} J. Eisert {\it et al.}, {\sl Phys. Rev. A} {\bf 62}, 52317 (2000).
\bibitem{huttonbose} A. Hutton and S. Bose, {\sl Phys. Rev. A} {\bf 69}, 042312 (2004).

\bibitem{plenio} J. Eisert {\it et al.}, {\sl Phys. Rev. Lett.} (to appear) (2004); M.B. Plenio and F.L. Semi{$\tilde{\mbox{a}}$}o, {\sl quant-ph/0407034}; M.B. Plenio {\it et al.}, {\sl New J. Phys.} {\bf 6}, 36 (2004).

\bibitem{myungBS} M.S. Kim {\it et al.}, {\sl Phys. Rev. A} {\bf 65}, 032323 (2002).

\bibitem{hillery} M. Hilllery {\it et al.}, {\sl Phys. Rev. A} {\bf 59}, 1829 (1999).

\bibitem{clone} G. De Chiara {\it et al.}, {\sl Phys. Rev. A} (to appear) (2004).

\bibitem{concurrence} W.K. Wootters, {\sl Phys Rev. Lett.} {\bf 80}, 2245 (1998).

\bibitem{npt}  A. Peres, {\sl Phys. Rev. Lett.} {\bf 77}, 1413 (1996); M. Horodecki, {\it et al.}, {\sl Phys. Lett. A} {\bf 223}, 1 (1996).

\bibitem{buzekimoto} M. Koashi, {\it et al.}, {\sl Phys. Rev. A} {\bf 62}, 050302 (2000).

\bibitem{cluster} H.J. Briegel and R. Raussendorf, {\sl Phys. Rev. Lett.} {\bf 86}, 910 (2001).




\bibitem{simon} R. Simon, {\sl Phys. Rev. Lett.} {\bf 84}, 2726 (2000).

\bibitem{myungmunro} M.S. Kim {\it et al.}, {\sl Phys. Rev. A} {\bf 66}, R030301 (2002).

\bibitem{referee} Z. Y. Ou, {\it et al.}, {\sl Phys. Rev. Lett.} {\bf 68}, 3663 (1992); Z.Y. Zou {\it et al.}, {\sl Appl. Phys. B} {\bf 55} 265 (1992).


\bibitem{polzik} C. Schori {\it et al.}, {\sl Phys. Rev. Lett.} {\bf 89}, 057903 (2002).



\bibitem{ww} R. Werner and M. Wolf, {\sl Phys. Rev. Lett.} {\bf 86}, 3658 (2001).

\bibitem{spiegazione} This is a consequence of the inequality $\hat{B}_{ab}(\varepsilon)\hat{S}_{a}\ket{00}_{ab}\neq\hat{S}_{a}\hat{S}_{b}\hat{S}_{ab}\ket{00}_{ab}$ holding for $\varepsilon\neq{\pi/4}$. $\hat{S}_{a}$ denotes single-mode squeezing. 

\bibitem{wvc} M.M. Wolf, {\it et al.}, {\sl Phys. Rev. Lett.} {\bf 92}, 087903 (2004).

\bibitem{buks} E. Buks and M.L. Roukes, {\sl JMEMS} {\bf 6}, 1057 (2002).


\bibitem{knobel} R.G. Knobel {\it et al.}, {\sl Nature} (London) {\bf 424}, 291 (2003).

\bibitem{armour} A.D. Armour {\it et al.}, {\sl Phys. Rev. Lett.} {\bf 88}, 148301 (2002).


\bibitem{zoller} I. Martin {\it et al.}, {\sl Phys. Rev. B} {\bf 69}, 125339 (2004).



\end{thebibliography}
\end{document}